\documentclass[
    ,final            % use final for the camera ready runs
  ]
  {aipproc}

\layoutstyle{8x11double}

\begin{document}

\title{The Comparison of the Swift Gamma-Ray Bursts With and Without Measured Redshifts}

\classification{01.30.Cc, 95.55.Ka, 95.85.Pw, 98.70.Rz}
\keywords      {gamma-ray astrophysics, gamma-ray bursts}

\author{David Huja}{address={Astronomical Institute of the Charles University, V Hole\v{s}ovi\v{c}k\'{a}ch 2, Prague, Czech Republic}}

\begin{abstract}

Gamma-ray bursts, detected by the Swift satellite, are separated into two samples: the bursts with and without determined redshifts. These two samples are compared by the standard Student t-test and F-test. We have compared the dispersions and the mean values of the durations, peak fluxes and fluences in order to find any differences among these two samples. No essential differences were found.

\end{abstract}

\maketitle

\section{The Sample}

We define two samples from the Swift dataset \cite{1}: The sample of GRBs without measured redshift $z$ (189 GRBs, 182 with measured duration $T_{90}$), and the sample with measured $z$ (97 GRBs, 94 with measured $T_{90}$). We know the name of GRB, its BAT duration $T_{90}$, BAT fluence at range 15-150~keV, BAT peak flux at range 15-150~keV, and redshift. Total number of the GRB studied is 286. The sample covers the period November 2004 - December 2007; the first event is GRB041227, the last one is GRB071227.

\section{Comparison of the peak flux}

The first sample without measured redshift contains 186 GRBs  with the mean value of the peak flux 2.7~ph$/$(cm$^{2}$s) and the dispersion 4.9~ph$/$(cm$^{2}$s). The second sample with measured redshift contains 95 GRBs, with the mean value of the peak flux 3.9~ph$/$(cm$^{2}$s) and the dispersion 6.3~ph$/$(cm$^{2}$s). The F-test for the dispersions tells $F$ = 1.66 (it belongs to the critical interval = $\langle 1.29; \infty)$), and thus the dispersions are not the same on the $\alpha$ = 5\%  level of the significance, \cite{3,4}. The Student t-test for the mean values gives t~=~1.605 (it does not belong to the critical interval = $\langle 1.645; \infty)$), and then the mean values of the peak fluxes are the same on average, \cite{3,4}. See Fig. 1.

\begin{figure}
  \includegraphics[width=0.8\textwidth]{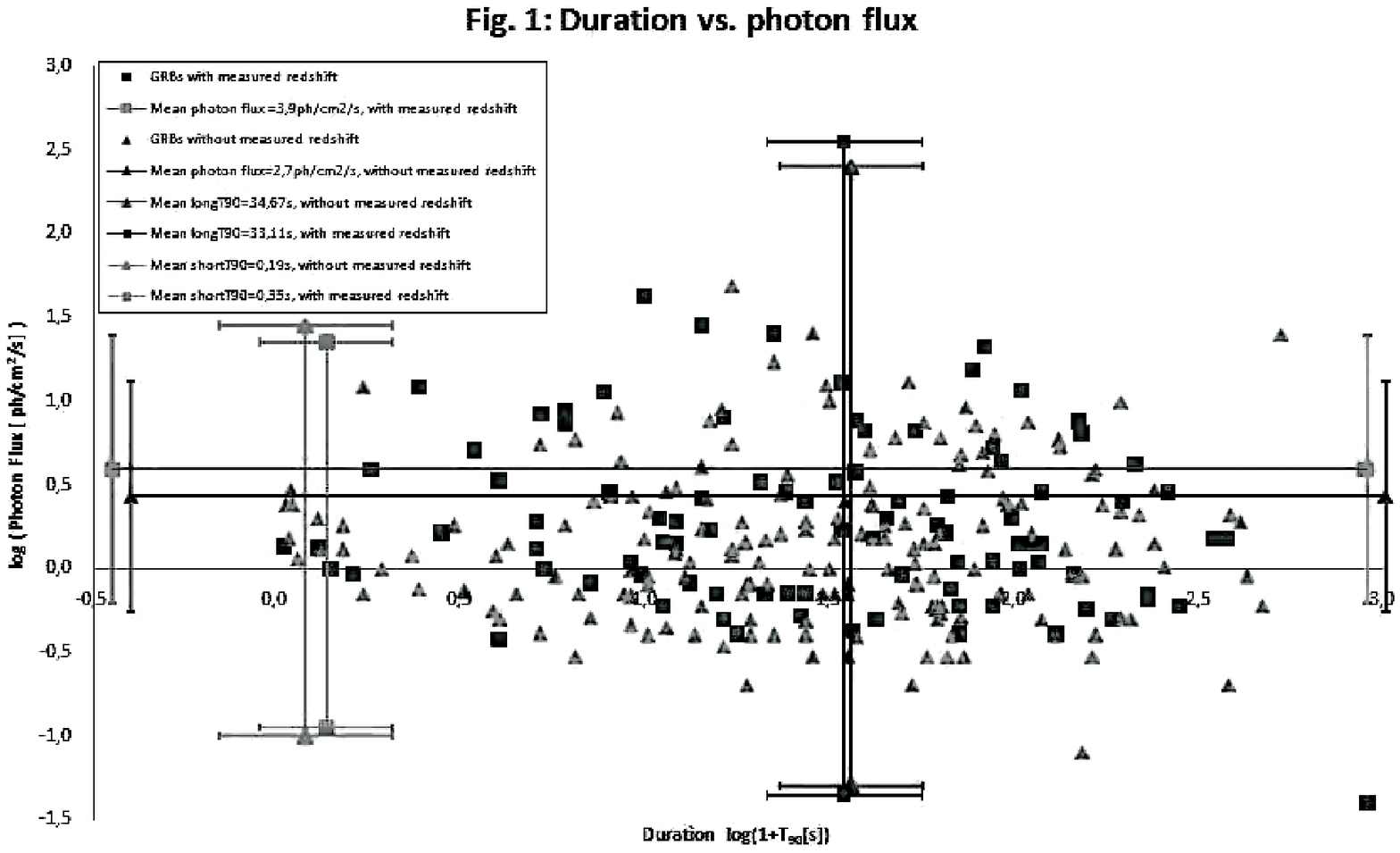}
\end{figure}

\section{Comparison of the fluence}

The first sample without measured redshifts contains 188 GRBs, with the mean fluence $29.4\times10^{-7}$~erg$/$cm$^{2}$ with the dispersion $80.3\times10^{-7}$~erg$/$cm$^{2}$. The second sample with measured redshifts contains 96 GRBs, with the mean peak flux $33.2\times10^{-7}$~erg$/$cm$^{2}$ with the dispersion $59.5\times10^{-7}$~erg$/$cm$^{2}$.

The F-test for the dispersions tells $F$ = 1.82 (it belongs to the critical interval = $\langle 1.29; \infty)$), hence the dispersions are not the same on average on the $\alpha$ = 5\% level of the significance, \cite{3,4}. The Student t-test for the mean values tells $t$ = -0.453 (it does not belong to the critical interval = $( -\infty; -1.645\rangle$) and then the mean fluences are the same on average \cite{3,4}. See Fig. 2.

\begin{figure}
  \includegraphics[width=0.8\textwidth]{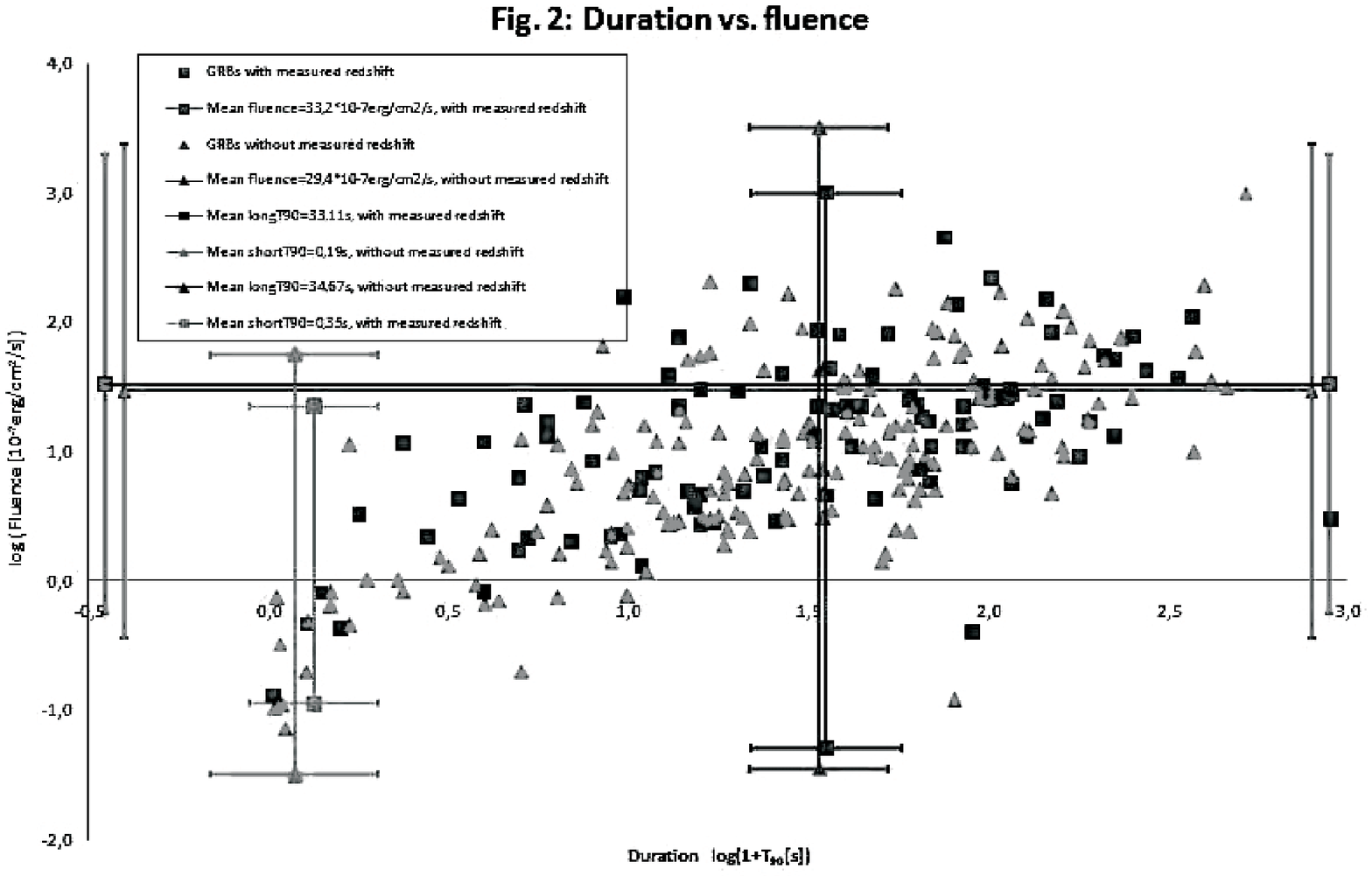}
\end{figure}

\section{Comparison of the duration $T_{90}$}

Having dozens of GRBs with measured redshifts  we can compare the two Swift samples with the standard F-test and Student t-test \cite{3,4}, for the sample without (182 GRBs) measured redshifts one obtained: $\mu_{1}$ = -0.73 ($T_{90}$(1) = 0.19s), $\sigma_{1}$ = 0.71, $\mu_{2}$ = 1.54 ($T_{90}$(2) = 34.67s), $\sigma_{2}$ = 0.56, the rate "short GRBs : long GRBs" = 9 : 91 (in \%, also absolutely 16 : 166); for the sample with (94 GRBs) measured redshifts one obtained: $\mu_{1}$ = -0.46 ($T_{90}$(1) = 0.35s), $\sigma_{1}$ = 0.51, $\mu_{2}$ = 1.52 ($T_{90}$(2) = 33.11s), $\sigma_{2}$ = 0.62, the rate "short GRBs : long GRBs" = 6 : 94 (in \%, also absolutely 6 : 88).

We have tested the rates of the long GRBs in the both samples. The t-test tells $t$ = -0.871 (it does not belong to the critical interval = $( -\infty; -1.645\rangle$), hence the rates of the long GRBs (and the short ones as well) are the same on average \cite{3,4}.

We have tested the dispersions with the F-test for the short GRBs. The $F$ = 1.938 (it does not belong to the critical interval = $\langle 4.619; \infty)$), the dispersions are the same on average. Then we have tested the mean $T_{90}$ for the short GRBs. The t-test tells $t$ = -0.975 (it does not belong to the critical interval = $( -\infty; -1.645\rangle$), hence the means are the same on average \cite{3,4}. We have tested the dispersions with the F-test for the long GRBs. The $F$ = 1.226 (it does not belong to the critical interval = $\langle 1.439; \infty)$), and thus the dispersions are the same on average. We have tested the mean $T_{90}$ for the long GRBs. The t-test tells $t$ = 0.253 (it does not belong to the critical  interval = $\langle 1.645; \infty)$), and then the means are the same on average \cite{3,4}.

\section{Conclusion}

The Swift satellite detected 189 GRBs (about 2/3 of the whole sample) without measured redshifts and 97 GRBs (about 1/3 of the whole sample) with measured redshifts. We compared these two samples in the mean values of the duration, peak fluxes and fluences, respectively. We have used  the F-test \cite{3,4} for comparison of their dispersions, and the Student t-test \cite{3,4} for comparison of  their mean values. We have determined that the mean values of all characteristics of the Swift GRBs are the same on average at the  $\alpha$ = 5\%  level of  the significance. But this is not the case for the dispersions. Hence, some differences exist in the two groups in accordance with \cite{5}, which can have an impact on the redshift distribution of GRBs \cite{6,7}. In our opinion the difference in the dispersions can be an instrumental effect, hence no essential difference is found between the two samples.

\begin{theacknowledgments}

This study was supported by the GAUK grant No. 46307, by the Grant Agency of the Czech Republic, grants No. 205/08/H005, and by the Research program MSM0021620860 of the Ministry of Education of the Czech Republic. The author appreciates valuable discussions and help of A. M\'esz\'aros and J. \v{R}\'{\i}pa.

\end{theacknowledgments}

\end{document}